\documentclass[aps,prl,showpacs,revsymb,amsfonts,amssymb,twocolumn,
groupedaddress,footinbib,floatfix]{revtex4-1}

\usepackage{graphicx}
\usepackage{amsfonts}
\usepackage{amssymb}
\usepackage{amsmath}
\usepackage{verbatim}
\usepackage{yfonts}

\usepackage[pdftex,bookmarks,colorlinks,breaklinks]{hyperref}  
\hypersetup{linkcolor=blue,citecolor=red,filecolor=green,urlcolor=green} 

\newcommand{\mbf}[1]{{\bf #1}}

\begin{document}

\title{Dynamic Aspects of Strong Pinning}

\author{A.U.\ Thomann}
\author{V.B.\ Geshkenbein}
\author{G.\ Blatter}
\affiliation{Institute for Theoretical Physics, ETH Zurich, 
8093 Zurich, Switzerland} 

\date{\today}

\begin{abstract}
We determine the current--voltage characteristic of type II superconductors in
the presence of strong pinning centers. Focusing on a small density of
defects, we derive a generic form for the characteristic with a linear
flux-flow branch shifted by the critical current (excess-current
characteristic).  The details near onset, a hysteretic jump (for $\kappa \gg
1$) or a smooth velocity turn-on ($\kappa \to 1$), depend on the
Labusch parameter $\kappa$ characterising the pinning centers.  Pushing the
single-pin analysis into the weak pinning domain, we reproduce the collective
pinning results for the critical current.
\end{abstract}

\pacs{74.25.F-, 74.25.Wx, 74.25.Sv}

\maketitle


The defining property of a (type II) superconductor is its ability to carry
electric current without dissipation. This superflow is destroyed  when the
magnetic induction $B$ enters the material in the form of quantized flux lines
or vortices \cite{abrikosov_57}: driven by the current density $j$ via the
Lorentz force $F_{\rm \scriptscriptstyle L} = j B/c$, the finite velocity $v$
of vortices generates a dissipating electric field $E = v B/c$ parallel to $j$
\cite{bardeen_65}.  It is the material defects immobilizing vortices which
reestablish the superflow of current, eventually rendering the superconductor
amenable to technological applications.  An elementary distinction is made in
the design and action of pinning defects: {\it strong} pins act individually
and generate large (plastic) deformations and metastable vortex states, while
{\it weak} defects are unable to pin vortices alone and thus act collectively.
In this letter, we determine the generic force--velocity (or current--voltage)
characteristic of vortices driven by a current $j$ and subject to a small
density $n_p$ of strong pins.

Vortex pinning has originally been studied by Labusch \cite{labusch_69} for
strong pins (see also  Ref.\ \onlinecite{campbell_72}) and has later been
extended to weak collective pinning by Larkin and Ovchinnikov
\cite{larkin_79}.  While the latter has been profoundly studied
\cite{blatter_94,brandt_nat_95_00}, the further development of strong pinning
theory has been less dynamic, although some progress has been made over time
\cite{campbell_78,matsushita_79,larkin_86,ivlev_91, koshelev_11}.  Recently,
the two regimes have been analyzed within a pinning diagram \cite{blatter_04}
delineating the origin of static critical forces $F_c$ as a function of defect
density $n_p$ and strength $f_p$. Here, we go beyond the calculation of the
static critical force $F_c$ and determine the full force $F_{\rm
\scriptscriptstyle L}\,( = j B/c)$ versus velocity $v\,( = c E/B)$ (or
$j$--$E$) characteristic of a so-called `hard' type II superconductor. We
focus on the single-pin--single-vortex strong pinning regime, implying that
defects are dilute and moderately strong, pinning only one vortex line at a
time; furthermore, we concentrate on isotropic material and ignore effects of
thermal fluctuations.
\begin{figure}[b]
\includegraphics[width=8cm]{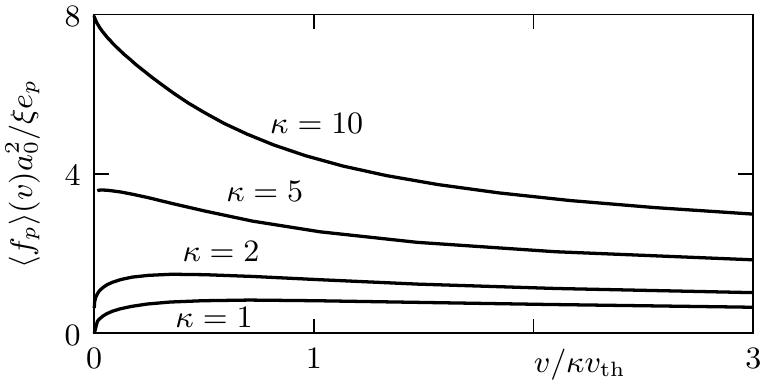}
\caption{ \label{fig:fp_Lorentz}
   Average pinning force $\langle f_p\rangle$ for a Lorentzian shaped pinning
   potential of various strengths. For strong pinning $\kappa \gg 1$ the
   critical force $f_c$ is large and the pinning force decays monotonously. On
   approaching the Labusch point $\kappa \to 1$ the critical force $f_c$
   vanishes and the pinning force is non-monotonic, first increasing $\propto
   \sqrt{v}$ and then decaying $\propto 1/\sqrt{v}$.}
\end{figure}

The calculation of critical forces for weak pins involves dimensional
\cite{larkin_79,blatter_94} or perturbative \cite{schmid_73,larkin_74}
estimates and is rather on a qualitative level. Calculations of the
force--velocity characteristic focus either on the perturbative regime at high
velocities \cite{schmid_73,larkin_74} or on the universal regime near
depinning \cite{nar-chauve_92_00}. The situation is different for strong
pinning: here, the critical force and the full dynamical response can be
determined quantitatively, once the shape of the pinning potential is known.
The force--velocity characteristic we find agrees well with numerous (even
textbook \cite{strnad_kim_64_65,huebener_79}) experimental results
\cite{berghuis_93,zeldov_02,pace_04}: a nearly linear flux-flow curve shifted
by the critical force $F_c$ (excess-current characteristic), with a hysteretic
jump in velocity at onset for strong pinning changing to a smooth rise on
approaching the weak pinning domain. Quite remarkably, continuing our
single-pin analysis into the weak pinning domain, we can find the usual weak
collective pinning results for the critical force. Below, we derive the
formalism leading us to the force--velocity characteristic, present the
results for the average pinning force $\langle F_p\rangle (v)$ for a
Lorentzian-shaped pin, derive the generic characteristic for the strong
pinning case in the dilute-pin limit, and finish with a rederivation of the
weak collective pinning results for the critical current from a study of
single-defect pinning.

The velocity--force characteristic derives from the dynamical equation for
vortex motion
\begin{equation}\label{eq:eom}
  \eta v = F_{\rm\scriptscriptstyle L}(j)-\langle F_p\rangle (v)
\end{equation}
with the Bardeen-Stephen \cite{bardeen_65} viscosity $\eta \sim B H_{c2}/
\rho_n c^2$ ($H_{c2}$ is the upper critical field and $\rho_n$ denotes the
normal state resistivity) and the velocity dependent average pinning force
density $\langle F_p\rangle(v)$ (we choose $\langle F_p \rangle(v)$ to be
positive). The pinning force density is determined by the positions $({\bf
R}_\mu + {\bf u}_\mu(z,t),z)$ of the flux lines (we choose ${\bf r}_\mu =
({\bf R}_\mu,z)$ to be the static lattice positions and ${\bf u}_\mu(z,t)$ is
the vortex displacement field; the vortex density is $a_0^{-2} = B/\Phi_0$
with $\Phi_0=hc/2e$ the flux quantum) and the individual pinning forces ${\bf
f}_p({\bf R} - {\bf R}_i) \delta(z - z_i)$ of defects located at positions
${\bf r}_{i} = ({\bf R}_i,z_i)$,
\begin{eqnarray}
   \label{eq:F_pin}
   \langle {\bf F}_p \rangle 
   &=&\frac{1}{N}\sum_\mu^N\int\frac{dz}{L}\, 
   \bf{F}_p({\bf r}_\mu,{\bf u}_\mu), \quad \textrm{with} \\
   \nonumber
   {\bf{F}}_p({\bf r}_\mu,{\bf u}_\mu)
   &=& \frac{-1}{a_0^2}
   \sum_i {\bf f}_p\bigl[{\bf R}_\mu+{\bf u}_\mu(z,t)-{\bf R}_i\bigr]
   \delta(z-z_i).
\end{eqnarray}
For a point-like defect $-e_p \delta({\bf r})$, the convolution with the
vortex shape $1-|\Psi({\bf r})|^2 \approx 2\xi^2/(R^2 + 2 \xi^2)$ ($\Psi$
denotes the complex order parameter and $\xi$ is the coherence length)
provides the pinning potential
\begin{eqnarray}\label{eq:varepsilon_p}
   \varepsilon_p({\bf R},z) = - e_p \frac{2 \xi^2}{R^2 + 2 \xi^2} \delta(z)
   \equiv e_p({\bf R}) \,\delta(z),
\end{eqnarray}
which here is of Lorentzian shape but may have another form in general. The
pinning force is given by the gradient ${\bf f}_p({\bf R}) = -\nabla_{{\bf
\scriptscriptstyle R}}e_p({\bf R})$.

The calculation of the pinning force density requires knowledge of the
displacement field ${\bf u}_\mu(z,t)$.  The latter is determined by the
solution of the dynamical equation which we write in integral form
\begin{eqnarray} \nonumber
   {\bf u}_{\nu}(z,t)
   &=& {\bf v} t + a_0^2\sum_{\mu} \int dz' dt' \, 
   \hat{\bf G}({\bf R}_\nu-{\bf R}_\mu, z-z', t-t')\\
   \noalign{\vspace{-5pt}}
   && \qquad\qquad\qquad\qquad\quad
   \times {\bf F}_p[{\bf r}'_{\mu},{\bf u}_\mu(z',t')].
   \label{eq:u_nu}
\end{eqnarray}
The first term accounts for the Lorentz force in Eq.\ (\ref{eq:eom})
generating the flux-flow velocity $v = F_{\rm\scriptscriptstyle L}/\eta$ in
the absence of pinning. The dynamical elastic Green's function $\hat{\bf
G}({\bf r},t)$ is given by the Fourier transform of the matrix
\begin{eqnarray} \label{eq:G}
   G_{\alpha\beta}(\mbf{k},\omega)&=&\frac{K_\alpha K_\beta/K^2}{c_{11}
   K^2+c_{44} k_z^2-i\eta\omega}\\ 
   && \nonumber \qquad\qquad\qquad 
   +\frac{\delta_{\alpha\beta} -K_\alpha
   K_\beta/K^2}{c_{66} K^2+c_{44} k_z^2-i\eta\omega}
\end{eqnarray}
with the elastic moduli $c_{11}$ (compression), $c_{44}$ (tilt), and $c_{66}$
(shear) \cite{blatter_94}.  The task simplifies considerably in the dilute-pin
limit (to order $n_p$) and for moderately strong pinning defects
trapping no more than one vortex; in this situation we can drop the sums over
$i$ and $\mu$ in Eqs.\ (\ref{eq:F_pin}) and (\ref{eq:u_nu}). We choose the pin
position at the origin and let the vortex move on the $x$-axis; the problem
then reduces to the calculation of the displacement field $u_x(z=0, t)$ at
$z=0$. With the asymptotic position $x=vt$ (at $z=\pm\infty$), we have to
solve the self-consistent equation
\begin{equation} 
   u(x) = x + \int_{-\infty}^x \frac{dx'}{v} \, G[0, (x-x')/v] \, f_p[u(x')],
   \label{eq:u}
\end{equation}
where $u(x) = u_x(z=0, t)$ and with $G = G_{xx}$~and $f_p$ the force along
$x$. Inserting the solution back into Eq.\ (\ref{eq:F_pin}), we obtain the
pinning force density $F_p(0,z,u) = -f_p[u(vt)] \, \delta(z)$. The
average pinning force density $n_p \, a_0^2 \int dz \langle F_p
\rangle$ due to a finite density $n_p$ of defects involves the average
$\langle \cdot\rangle$ over pin locations and time; the latter transform to
an average along $x$ and the impact parameter $b$ of the vortex on the defect,
\begin{equation}
   \langle F_p\rangle(v) = n_p \langle f_p \rangle
   = - n_p \Bigl\langle \int_{-\infty}^\infty \frac{dx}{a_\parallel}
   f_p[u(x)]\Bigr\rangle_b,
   \label{eq:av_F}
\end{equation}
where $a_\parallel$ is the distance between vortices along the $x$ direction.
Restricting ourselves to the case of strongest pinning with $b=0$ and treating
all trajectories within the range $\sigma\sim\xi$ of the pin equally, the
average over impact parameters $\langle \cdot \rangle_b$ contributes a factor
$\sigma/a_\perp$ with $a_\perp$ the transverse distance to the next vortex,
hence $a_\parallel a_\perp = a_0^2$. Inserting the result for $\langle
F_p\rangle(v)$ back into the dynamical equation (\ref{eq:eom}) and solving for
the velocity $v$ for a given current $j$ provides us with the desired result,
the force--velocity characteristic of the superconductor.

In the static situation, the self-consistent integral equation (\ref{eq:u})
simplifies to the algebraic equation
\begin{equation}\label{eq:u_s}
   u_s(x)= x +{f_p[u_s(x)]}/{\bar{C}}, 
\end{equation}
with $\bar{C}^{-1}= G({\bf r} = 0, \omega = 0)$ the local static elastic
Green's function, $\bar{C} \sim \varepsilon_0/a_0$ with $\varepsilon_0 =
(\Phi_0/4\pi \lambda)^2$ the energy scale for vortices. Strong pinning is
characterized by the appearance of bistable solutions in Eq.\ (\ref{eq:u_s}),
implying that the derivative
\begin{equation}\label{eq:u_s^p}
   \frac{d}{dx} \, f_p[u_s(x)]= -\bar{C}
   \frac{f_p^\prime[u_s(x)]}{f_p^\prime[u_s(x)]-\bar{C}} 
\end{equation}
of the effective force has to diverge---this provides us with the Labusch
criterion \cite{labusch_69} $\kappa \equiv \max_x\bigl\{
f_p^\prime[u_s(x)]/\bar{C}\bigr\} = 1$ separating weak ($\kappa <1$) and
strong ($\kappa > 1$) pinning. Note that the effective force gradient inside a
very strong pin is universally given by the effective elastic constant
$\bar{C}$, not by $f_p'$. The different solutions of Eq.\ (\ref{eq:u_s}) at
$\kappa > 1$ are associated with the unpinned ($u_s$ outside the pin) and
pinned ($u_s$ inside the pin) states of the vortex; their asymmetric
statistical occupation at finite drive produces a finite critical force
density $F_c = \max \langle F_p \rangle(v=0)$ where the maximum is taken over
the pinned and unpinned branches. For a weak pin ($\kappa < 1$), Eq.\
(\ref{eq:u_s}) has a unique solution and the critical force density $F_c$
vanishes.
\begin{figure}[b]
\includegraphics[width=8.0cm]{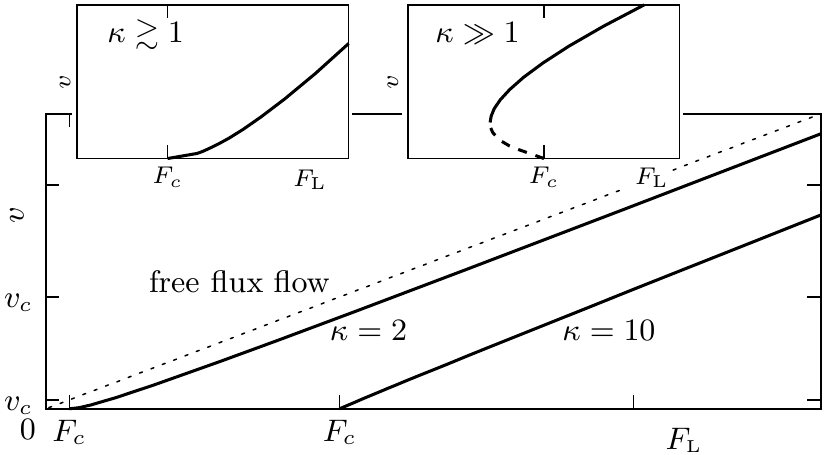}
\caption{ \label{fig:fv}
   Force-velocity characteristic for a Lorentzian-shaped pinning potential.
   For small defect densities $n_p a_0 \xi^2 \kappa = 0.05$ we find an
   excess-current characteristic, a shifted (by $F_c \propto n_p$) linear
   curve with a slope reflecting flux-flow behavior and approaching the true
   (unshifted) flux-flow behavior only at high velocities $\kappa
   v_\mathrm{th} \gg v_c$. The insets sketch the behavior near critical, with a
   hysteretic jump of order $v_\mathrm{nl} \propto n_p^2$ appearing at strong
   pinning $\kappa \gg 1$ and a smooth onset $v \propto
   v_\mathrm{th}(\kappa-1)^4 \, (F_{\rm \scriptscriptstyle L}/F_c-1)^2$ on
   approaching the Labusch point $\kappa \to 1$.  }
\end{figure}

In the dynamical situation with $v >0$ we have to solve the self-consistent
integral equation (\ref{eq:u}) and thus need to know the time dependence of
the Green's function $G(0,t)$. At short times $t < t_\mathrm{th} = \eta a_0^2
/ 4 \pi c_{66}$ the Green's function is dominated by the response of an
individual vortex line (the one-dimensional (1D) regime), $G^{\rm
\scriptscriptstyle 1D} (0,t>0) \sim (t_\mathrm{th}/t)^{1/2}/\bar{C}
t_\mathrm{th}$; at long times $t > t_\mathrm{th}\lambda^2/a_0^2$ the full 3D
vortex system provides the response and $G^{\rm \scriptscriptstyle 3D}(0,t>0)
\sim (t_\mathrm{th}/t)^{3/2}a_0/\bar{C} t_\mathrm{th}\lambda$ (the
intermediate dispersive or 4D regime with $G^{\rm \scriptscriptstyle 4D} \sim
(t_\mathrm{th}/t)^2/\bar{C} t_\mathrm{th}$ is less relevant in our analysis
below).  

At high velocities $v$ the time integral in Eq.\ (\ref{eq:u}) extends over
short times and the velocity-dependent part of the pinning force scales as $t
G^{\rm \scriptscriptstyle 1D} \propto \sqrt{t}$, while, at small velocities,
long times are relevant and $t G^{\rm \scriptscriptstyle 3D} \propto
\sqrt{1/t}$. The time $t$ to velocity $v$ transformation $t \sim
\sigma_\mathrm{eff}/v$ involves the effective pin size $\sigma_\mathrm{eff}
\sim \kappa \xi/ (1+v/\kappa v_\mathrm{th})$ which depends on the pinning
strength $\kappa$ and on the velocity $v$ itself \cite{thomann_11}; for
$\kappa \to 1$ and at high velocities $v > \kappa^2 v_\mathrm{th}$ the
effective pin size $\sigma_\mathrm{eff}$ saturates at the true geometric pin
size $\xi$ (here, $v_\mathrm{th} \sim \xi/t_\mathrm{th}$ is the basic velocity
scale).  The corrections to the critical force $F_c$ at small velocities $v <
(a_0^2/\lambda^2) \kappa v_\mathrm{th}$ then are expected to scale as
$\sqrt{v/\kappa v_\mathrm{th}}$, while the high-velocity $v > \kappa^2
v_\mathrm{th}$ corrections to the dissipative force $\eta v$ (flux-flow) decay
as $\sqrt{v_\mathrm{th}/v}$.  This is confirmed by the numerical solution of
the problem following the steps indicated above and where the results are
shown in Fig.\ \ref{fig:fp_Lorentz} (we assume non-dispersive moduli
corresponding to a field $B \sim \Phi_0/\lambda^2$).  The forward integration
of Eq.\ (\ref{eq:u}) has been done for a Lorentzian-shaped potential of the
form (\ref{eq:varepsilon_p}) and different pinning strengths as expressed by
the Labusch parameter $\kappa \sim (e_p/\xi \varepsilon_0) (a_0/\xi)$; with
$e_p \sim H_c^2 \xi^3 \sim \varepsilon_0 \xi$ ($H_c$ the thermodynamic
critical field) the Labusch parameter can naturally access large numbers
$\kappa \sim a_0/\xi \gg 1$. The scaled average pinning force $\langle
f_p\rangle a_0^2/e_p\xi$ is plotted against the scaled velocity $v/ \kappa
v_\mathrm{th}$ and exhibits a monotonic decrease at large $\kappa$ and a
non-monotonic behavior enforced by the vanishing of the critical force $f_c =
\langle f_p\rangle(0)$ as $\kappa \to 1$. While our rough estimate above
correctly predicts the shape $\propto \sqrt{v/\kappa v_\mathrm{th}}$ of the
finite-velocity corrections, its sign depends on $\kappa$ in a nontrivial way
\cite{thomann_11}.  Note that we plot the single-pin result $\langle
f_p\rangle$ rather than the corresponding force density $\langle F_p\rangle$
as the density $n_p$ is an important independent parameter.
%
%
%
%

In our discussion of the force--velocity characteristic we first concentrate
on the overall shape away from the onset of vortex motion. The generic
characteristic
\begin{equation}\label{eq:fv}
   \frac{F_{\rm\scriptscriptstyle L}}{F_c}= \frac{v}{v_c}
   + \frac{\langle f_p\rangle (v/\kappa v_\mathrm{th})}{f_c}
\end{equation}
involves two velocity scales, the velocity $\kappa v_\mathrm{th}$ governing
the pinning force $\langle f_p\rangle$ (as confirmed by a detailed analysis of
Eq.\ (\ref{eq:av_F}) \cite{thomann_11}) and the scale $v_c = F_c/\eta$
appearing from the competition between the dissipative ($\eta v$) and the
critical ($F_c = n_p f_c$) force densities.  In the limit of small pin
densities $n_p$, the linear term in Eq.\ (\ref{eq:fv}) changes on the small
velocity scale $v_c \propto n_p$, while the pinning force $\langle
f_p\rangle(v)$ deviates from its static value $f_c$ only on the larger scale
$\kappa v_\mathrm{th}$ which does not depend on $n_p$.  Indeed, making use of
the expression $F_c \sim (\xi^2/a_0^2) n_p f_p (\kappa-1)^2/\kappa$ for the
critical force density \cite{blatter_04}, we find the ratio $v_c/\kappa
v_\mathrm{th} \sim n_p a_0 \xi^2 (\kappa-1)^2/\kappa \ll 1$ in the small pin
density limit and at fixed $\kappa$ (note that the limit $\kappa \to 1$ at
fixed $n_p$ would take us out of the single-pin regime).  With $\langle
f_p\rangle(v) \approx f_c $ for velocities $v \sim v_c \ll \kappa
v_\mathrm{th}$ we find a characteristic that takes the generic form of a
shifted (by $F_c$) linear (flux-flow) curve, $v \approx (F_{\rm
\scriptscriptstyle L} - F_c)/\eta$, see Fig.\ \ref{fig:fv}; the free
dissipative flow $v = F_{\rm \scriptscriptstyle L}/\eta$ is approached only at
very high velocities $v \gg \kappa v_\mathrm{th} \gg v_c$.  The simple
excess-current characteristic is a consequence of the separation of velocity
scales $v_c$ and $\kappa v_\mathrm{th}$; the latter merge at strong pinning
with increasing density $n_p$ when strong 3D pinning goes over into 1D strong
pinning at $n_p a_0 \xi^2 \kappa \sim 1$~\cite{blatter_04}. Using qualitative
arguments, a similar excess current characteristic has been found in Ref.\
\onlinecite{campbell_78}.

The above simple overall structure of the force--velocity characteristic is
modified at very small velocities and in close vicinity to the critical force
density $F_c$; in this regime we can rewrite Eq.\ (\ref{eq:fv}) in the simple
form $F_{\rm\scriptscriptstyle L}/F_c= v/v_c +1 \pm (v/v_p^\pm)^{1/2}$, where
the `$+$' (`$-$') sign applies to the limits $\kappa \to 1$ ($\kappa \gg 1$).
The small-velocity pinning scales $v_p^\pm$ derive from the 3D expression of
the pinning force density \cite{thomann_11} $\langle F_p\rangle -F_c \sim
(\xi^2/a_0\lambda) n_p f_p \kappa \sqrt{v/\kappa v_\mathrm{th}}$ (at large
$\kappa$), $v_p^- \sim (\lambda^2/a_0^2)\kappa \,v_\mathrm{th}$ for $\kappa
\gg 1$ and $v_p^+ \sim (\lambda^2/a_0^2) (\kappa-1)^4\, v_\mathrm{th}$ for
$\kappa \to 1$.  For strong pinning, the negative (non-linear) correction in
the average pinning force density generates a bistability (and hence
hysteretic jumps) on the scale $v_\mathrm{nl} \sim v_c^2/v_p^- \propto n_p^2$.
On the other hand, approaching the Labusch point, the correction changes sign
and the velocity increases quadratically $v \sim v_p^+ (F_{\rm
\scriptscriptstyle L}/F_c-1)^2$ until crossing over into the linear regime at
$v_\mathrm{nl} \sim v_c^2/v_p^+ \ll v_c \ll v_p^+$. These features are visible
in the insets of Fig.\ \ref{fig:fv} showing an expanded view of the
characteristic near onset.

Next, we push our single-pin (SP) analysis into the weak pinning domain
$\kappa < 1$ and establish its relation to weak collective pinning (WCP)
theory. In the dynamical formulation of WCP, we determine the pinning force
$\langle F_p\rangle^{\rm \scriptscriptstyle WCP} (v)$ perturbatively (to
lowest order in $\kappa$ and $n_p$) at high velocities and follow the velocity
correction $\delta v = \langle F_p\rangle^{\rm \scriptscriptstyle WCP}
(v)/\eta$ down to small $v$. As the correction $\delta v$ becomes of order
$v$, higher order terms become relevant \cite{larkin_74} and we stop the
analysis, interpreting the breakdown of perturbation theory as the signature
of a finite critical force $F_c^{\rm \scriptscriptstyle WCP}$. The latter then
derives from the critical velocity $v_c$ defined through the criterion
$\langle F_p\rangle^{\rm \scriptscriptstyle WCP} (v_c) \sim \eta v_c =
F_c^{\rm \scriptscriptstyle WCP}$.

Within the SP analysis valid at small densities $n_p$, we usually calculate the
pinning force density $\langle F_p\rangle^{\rm \scriptscriptstyle SP} (v)$
exactly, cf.\ Fig.\ \ref{fig:fp_Lorentz}; in the case of weak pinning $\kappa
\ll 1$, we can use perturbation theory as well and we find the result
\begin{equation}
   \langle F_p\rangle^{\rm \scriptscriptstyle SP}(v)
   \approx \int_0^\infty dt \, G(0,t)\, K^{x\alpha\alpha}(vt,0),
   \label{eq:av_F_pert}
\end{equation}
where we have expanded the average pinning force density Eq.\ (\ref{eq:av_F})
for a displacement $u(x) = x + \delta u(x)$ close to flux-flow and used the
lowest order (in $\kappa$) approximation of Eq.\ (\ref{eq:u}) for $\delta
u(x)$. In Eq.\ (\ref{eq:av_F_pert}), $K({\bf u})= (n_p/a_0^2) \int d^2 R \,
e_p({\bf R} - {\bf u})\, e_p({\bf R})$ replaces the usual pinning energy
correlator showing up in WCP theory \cite{blatter_94} (the superscripts denote
derivatives with respect to $u_x$ and $u_\alpha$). Hence, the corrections
$\delta v$ from both the WCP- and the SP analysis agree with one another to
lowest order in $\kappa$ and in the pin density $n_p$. The difference in the
two approaches arises when we take the velocity $v$ to zero: While we stop at
$\delta v \sim v$ and arrive at a finite $F_c^{\rm \scriptscriptstyle WCP}$ in
WCP, we take $v$ all the way to zero within the SP analysis and obtain a
vanishing critical force $F_c^{\rm \scriptscriptstyle SP} = 0$. On the other
hand, using the SP result Eq.\ (\ref{eq:av_F_pert}) and adopting the WCP
cutoff, we find a finite critical current as well: with the estimate $\langle
F_p\rangle^{\rm \scriptscriptstyle SP} (v) \sim n_p (\xi/\lambda)
(f_p^2/\varepsilon_0) (v/v_\mathrm{th})^{1/2}$ valid at low velocities and the
conditions $\langle F_p\rangle^{\rm \scriptscriptstyle SP}(v_c) \sim \eta v_c
\sim j_c B/c$, we obtain the critical current
\begin{equation}
   \label{eq:jc}
   j_c \sim j_0 (\xi^2/\lambda)^2
   \bigl(n_p a_0^3 f_p^2/\epsilon_0^2\bigr)^{2} \propto n_p^2,
\end{equation}
in agreement with the results obtained from weak collective pinning theory
\cite{blatter_04}. This result is quite remarkable: first, the critical
current (\ref{eq:jc}) is proportional to $n_p^2$, the {\it square} of the pin
density $n_p$, {\it i.e.}, its origin is in the correlations between pins.
Second, the result is still consistent with the standard SP result $\langle
F_p\rangle^{\rm \scriptscriptstyle SP} (v=0) = 0$, as the latter is an order
$n_p$ result and corrections $\propto n_p^2$ are beyond the standard SP
approach. Going back to strong pinning $\kappa >1$, we already obtain a finite
critical force $\langle F_p\rangle^{\rm \scriptscriptstyle SP}(v=0) \propto
n_p$, {\it linear} in pin density. Pin-pin correlations then are expected to
provide corrections $o(n_p)$ which vanish faster than linear and we can
approach the critical force parametrically closer than in the WCP case.
 
Comparing our theoretical results to typical measured current--voltage
characteristics, we find good agreement with experimental results
\cite{strnad_kim_64_65,huebener_79,berghuis_93,zeldov_02,pace_04}. The
excess-current characteristic reported in these experiments was pointed out
early on by Campbell and Evetts\cite{campbell_72}, however, we are not aware
of any `microscopic' derivation of this basic result.  Unfortunately, a
detailed comparison between theory and experiment is still not available
today. Given a specific material, the defect structure is usually non-trivial
and may include a variety of pin types.  Furthermore, the parameters
characterizing the defects are difficult to find.  Experiments with
superconductors where defects could be designed, tuned, and properly
characterized would provide a great help and motivation in further developing
the theory of pinning, particularly the crossover regime between strong and
weak collective manifesting itself first in the small-velocity domain.

With thankfulness and in memoriam of Anatoli Larkin who has initiated this
study. We acknowledge financial support of the Fonds National Suisse through
the NCCR MaNEP.

\end{document}